\documentclass[natbib]{emulateapj}
\usepackage[]{natbib}

\newcommand{\mjup}{M$_{Jup}$}

\newcommand{\rjup}{R$_{Jup}$}
\newcommand{\teff}{T$_{\rm eff}$}
\newcommand{\lprime}{${\rm L}^{\prime}$}
\newcommand{\mlprime}{M$_{\rm L}^{\prime}$}
\newcommand{\logg}{log($g$)}
\newcommand{\mearth}{M$_{Earth}$}

\newcommand{\kms}{km/s}

\slugcomment{Accepted to ApJ}
\shorttitle{IR imaging of HR8799}
\shortauthors{Hinz et al.}
\begin{document}

\title{Thermal Infrared MMTAO Observations of the HR 8799 Planetary System
\footnotemark[1]}

\footnotetext[1]{Observations reported here were obtained at the MMT Observatory, 
a joint facility of the University of Arizona and the Smithsonian Institution.}

\author{Philip M. Hinz \footnotemark[2] $^{,}$\footnotemark[3], 
Timothy J. Rodigas\footnotemark[2], 
Matthew A. Kenworthy\footnotemark[2], 
Suresh Sivanandam\footnotemark[2], 
Aren N. Heinze\footnotemark[4], 
Eric E. Mamajek\footnotemark[5], 
Michael R. Meyer\footnotemark[2] $^{,}$\footnotemark[6]}

\footnotetext[2]{Steward Observatory, The University of Arizona, 933 N. Cherry Ave., 
Tucson, AZ 85721, USA}
\footnotetext[3]{phinz@as.arizona.edu}
\footnotetext[4]{Department of Physics and Astronomy, 
Swarthmore College, Swarthmore, PA 19081, USA}
\footnotetext[5]{ University of Rochester, 
Department of Physics \& Astronomy, Rochester, NY, 14627-0171, USA}
\footnotetext[6]{Institue for Astronomy, Swiss Federal Institute of Technology (ETH),
CH-8093 Zurich, Switzerland }

\begin{abstract} 

We present direct imaging observations at wavelengths of 3.3, 3.8 (\lprime\,band), 
and 4.8 (M band) \micron, for the planetary system surrounding HR 8799.  All three planets 
are detected at \lprime\,. The $c$ and $d$ component are detected 
at 3.3 \micron, and upper limits are derived from the M band observations. 
These observations provide useful constraints on warm giant planet atmospheres.  
We discuss the current age constraints on the HR 8799 system, and show that several potential
co-eval objects can be excluded from being co-moving with the star. 
  Comparison of the photometry is made to models for giant planet atmospheres.  
  Models which include non-equilibrium chemistry provide a reasonable match to  the colors
of $c$ and $d$.  From the observed colors in the thermal infrared we estimate \teff$<$ 960 K for $b$, and \teff=1300 and 1170 K
for $c$ and $d$, respectively.   This provides an independent check on the effective temperatures and thus masses of the objects from
the \citet{ Marois08} results.
\end{abstract}


\section{Introduction}
Over the past decade, a number of techniques have dramatically expanded our 
understanding of the nature of exoplanets. Initial detection of
systems via radial velocity variations has been followed up by
studies of transits, astrometric confirmation, and gravitational
microlensing.  All of these approaches have been helpful in developing
our picture of planetary system architectures, and providing insight into
formation.

The direct imaging of extrasolar planets is the latest technique to
provide useful information, with detection, of objects, first, around low-mass objects \citep{Chauvin04}, followed, 
more recently, by the detection of several planets around intermediate mass stars \citep{Kalas08, Marois08, Lagrange08}.  

Direct images of extrasolar planets not dominated by insolation have the potential to provide a
wealth of information about the size, temperature, composition, and
even formation history of these objects.  The recently discovered
planetary system around HR 8799 \citep{Marois08}, with three massive
planets at large orbital separations, may provide one of the most useful laboratories to constrain
the spectral energy distribution of young giant planets.

HR 8799 is a young A5V star, thought to be approximately 30-160 Myr old \citep{Marois06}.  
It is known to have a bright debris disk at large orbital radii
\citep{Williams06} with approximately 0.1 \mearth\, of material at
50\,K.  A tenuous inner disk is thought to exist as well
\citep{Chen06}.  Taken together with the planets, this suggests a
system with an inner dust disk truncated by the $d$ component at 24
AU, and an outer disk truncated by the $b$ component beyond $\sim$ 80
AU. Su et al. (2009) also find an outflow of small grains, likely caused by
gravitational stirring from the planets.  The star also has a number of other interesting
properties, including X-ray emission \citep{Hearty99, Schroeder07} and
a deficiency in refractory metals in the stellar atmosphere \citep[the $\lambda$ Boo
phenomenon;][]{Gray99, Gray03}. Multiple metallicity estimates are
consistent it being metal-poor 
\citep[Fe/H $\simeq$\, -0.5;][]{Gray99}, but having solar abundance
 lighter elements, such as carbon and oxygen \citep{Sadakane06, Gerbaldi07}. 

The planets around HR 8799 are interesting, particularly in the context
of planet formation alternatives.  If the planets formed in situ, the
core accretion hypothesis \citep{Pollack96} would require formation of
10 \mearth\, cores at distances of 40 and 70 AU well before the
dispersal of the gas.  A disk fragmentation scenario \citep{Boss97, Nero09, Dodson09, Helled09}
may be able to more naturally explain the massive planets and their
location.  A more detailed look at the planet's environment and their
spectral energy distributions (SEDs) can provide clues to
these alternatives.

Models of the spectral energy distribution for planets \citep{Burrows97,Baraffe03} are 
used to estimate the temperature and mass of the
objects.  However, these models are currently only constrained by
field brown dwarfs and objects such as 2M1207, whose formation history
is uncertain \citep{Lodato05, Mamajek07}.  Are giant planets formed in
a circumstellar disk around a normal star different?  Objects such as
the planets around HR 8799, as well as the planets around Fomalhaut
\citep{Kalas08} and $\beta$ Pic \citep{Lagrange08} may provide our first opportunity
to address this topic.  With multi-wavelength constraints of the
brightness of the planets we can disentangle effects from clouds,
composition, and vertical mixing that may affect the measurement of the 
temperature, and thus the mass of the planets.  By comparison of observations with 
field objects and current models, we may be able to develop a better 
understanding of the physics leading to the SED's of giant exoplanets.

The 3-5 \micron\, region provides access to both the 3.4 \micron\, CH$_4$ feature 
and the CO bandhead at 4.7 \micron.  Observations of this portion of the SEDs provides useful
constraints on the relative amount of CO to CH$_4$ in the atmospheres
of these planets.  The relative absorption of these species, in the near-infrared, is used to
define the transition in spectral type between the L and T sequence among field brown dwarfs. 

Observations of brown dwarfs in this region \citep{Leggett07}  indicate that their colors 
are best reproduced by models that have substantial vertical mixing between the hot lower 
layers and the cooler upper atmosphere. Comparison of HR 8799 planets to these objects
will allow us to better understand how closely exoplanet atmospheres may mimic those of 
field brown dwarfs. 

Constraining the SED of giant planets in this region may help guide future direct imaging planet searches. Both
Jupiter \citep{Gillett69} and Gliese 229 B \citep{Oppenheimer95} 
have a broad peak in the flux at 4-5 \micron, suggesting that this
may be a robust feature of brown dwarfs and giant planets in the \teff=100-1000 K 
range.  However, various models indicate different bands as being preferable (c.f. \citet{Marley96}, and  \citet{Burrows97}).
In order to understand whether searching in L' or M band is preferable,
it would be helpful to understand giant planet colors over the expected temperature range of these 
objects.

In this paper we present observations at 3.3, 3.8 and 4.8 \micron\, of
the planets orbiting HR 8799.  In Section 2 we describe the MMTAO and
Clio camera observations of the HR 8799 system.  Section 3 details the
data reduction and analysis. Section 4 describes the photometric and
astrometric results.  Section 5 analyzes the age estimates for HR 8799.  
Section 6 compares these results to theoretical
models and observations of brown dwarfs in the field.  Finally, in
Sections 7 and 8 we discuss the interpretation of the results and
summarize our conclusions. 

\section{Observations}

HR 8799 was observed on 21 November 2008,  8
January 2009, and 12 September 2009, using the MMT deformable secondary-based adaptive optics
system \citep{Wildi03, Brusa04}. Cirrus clouds intermittently affected
the throughput of the observations on 21 November 2008, and 12 September 2009.

The Clio camera \citep{Freed04, Sivanandam06}, an optimized instrument 
for exoplanet detection \citep{Hinz06, Heinze06} in this spectral
region, was used for the observations.  The camera has a high flux
detector that allows high duty cycle observations in the \lprime\,
and M band atmospheric windows.  Combined with the low background
provided by a deformable secondary \citep{Lloyd-Hart00}, the system is
capable of sensitive observations in the thermal infrared.

The Clio observations use the Barr M band ($\lambda$=4.8 \micron, $\Delta\lambda$=0.6 \micron, \citet{Simons02}), rather
than the more common MKO M' passband ($\lambda$=4.68 \micron, $\Delta\lambda$=0.22 \micron).  The Barr M band filter is 
preferred for sensitive M band observations that maximize the signal-to noise for the expected 
SED of a cool object \citep{Freed04}. At \lprime\, the MKO filter was used.
The zeropoints used for these filters are 248 Jy and 154 Jy for \lprime\, and M 
respectively \citep{Tokunaga05, Bessell88}. 

In addition to standard MKO  \lprime\, and Barr M band filters, we carried out
observations with a filter which was centered at 3.3\,\micron\, with
half-power points at 3.10 \micron\, and 3.5 \micron.  This filter is
well matched to the methane absorption located at 3.4 \micron\, and
lies in the short wavelength region of the L band atmospheric window.
However, it is not closely related to any standard photometric system
in the L window.  We refer to it in the rest of the paper as the 3.3
\micron\, filter, and list magnitudes for this filter with the notation [3.3].  The zeropoint adopted for this filter, is 330 Jy,
obtained by interpolating between the L' and K band photometric zeropoint.

Clio observations are obtained by acquiring individual integrations of
the detector with frame times set to avoid saturation by the
background emission, typically 0.2 s for M band, 1.5 s for \lprime\, band.
These individual exposures are accumulated into cubes of data with
approximately 15-20 seconds total integration. Typically, five
exposures are obtained before nodding the telescope in elevation.
This results in the star being moved approximately every minute from
one side of the detector to the other to provide periodic measurements
of the sky background, as well as detector variations.  Each nod pair
provides a complete set of data with ten independent measurements of
the field around HR 8799.

A significant source of noise in the images are static or slowly changing
speckles.  The effect of these are minimized with the technique of angular differential imaging
\citep[e.g.][]{Marois06, Hinz06}. During the observations the instrument derotator 
is not operated. This has the effect of causing the sky coordinates to slowly rotate
relative to the detector coordinates.    Since the static speckles stay fixed with 
respect to the telescope, we can remove them in the final image processing without
affecting the flux from an off-axis companion.  

\begin{figure}
\includegraphics[angle=0,scale=.6]{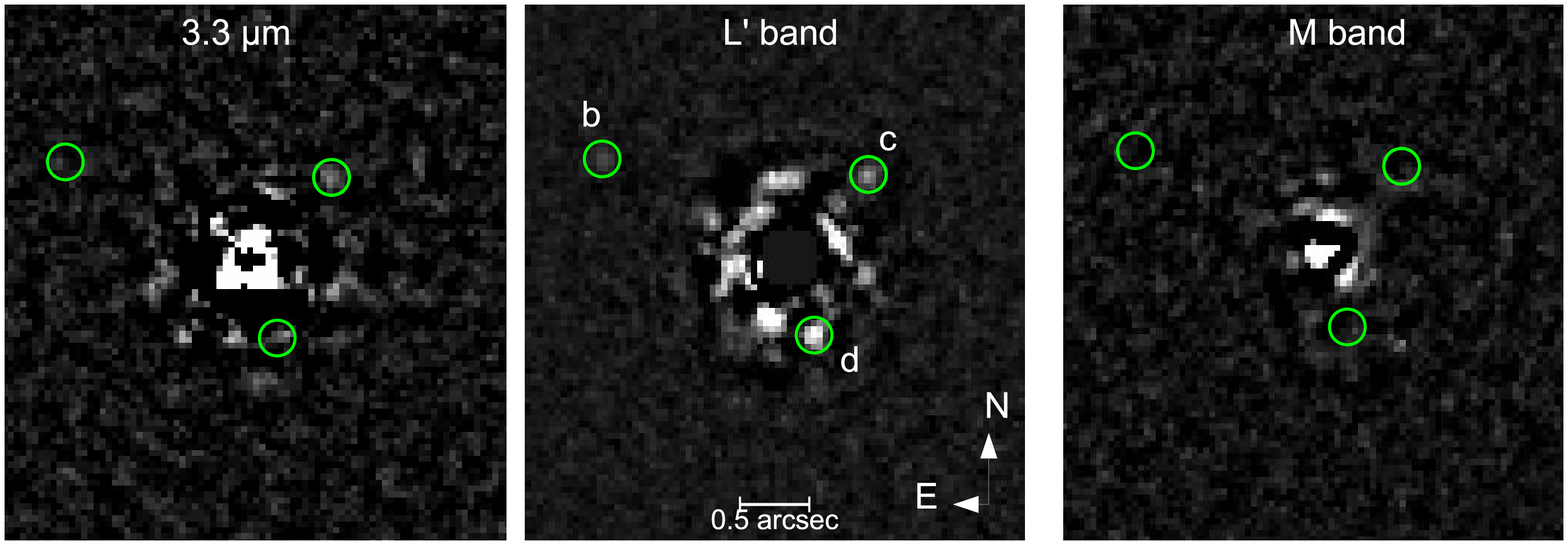}

\caption{Images at 3.3\micron, \lprime\, (3.8 \micron) and M (4.8 \micron) of 
HR 8799.  All three components are detected at \lprime\, (center). At 3.3
\micron(left) only $c$ and $d$ are detectable.  At M band (right) none of the
planets are detectable above the sky background noise.  The circles identifying the locations of
the planets in each image are the approximate size of the photometric aperture used.
\label{imfig}}
\end{figure}

Observations were taken in both \lprime\, and M band on November 21, 2008.
Approximately 70 degrees of rotation was obtained at \lprime\, and 30 degrees
at M band (see Table \ref{table1}).  For \lprime\, band observations, 30
frames were rejected due to image jitter while observing near
transit. The final image (see Figure \ref{imfig}) uses approximately
90 minutes of data.  The M band observations were taken just after
this, as intermittent cirrus clouds began to form.  Consequently,
although approximately 170 minutes of integration was acquired, only
84 minutes of total integration (210 total frames) were used, after
rejecting those images which showed stellar fluxes less than approximately 60\% of the "clear"
images.

A first set of 3.3\,\micron\, observations were taken on January 9 2009, which
required observing the star well after it had transited the meridian.
This resulted in only a small amount of sky rotation for these
observations.  Consequently, for the 3.3\,\micron\, observations we
observed a standard star for use in PSF subtraction.  A total of 48
minutes of integration was obtained in the 3.3\,\micron\, filter.

A second set of 3.3\,\micron\, observations were taken on 12 September 2009.
A total of 113 minutes of integration was obtained, with approximately 30 degrees
of rotation during the observations.  Light clouds were present throughout the observations.

As can be seen in Figure \ref{imfig}, slowly changing speckles have different
levels of importance for the three observed wavelengths.  At 3.3 and
3.8 \micron\, the noise in the residual subtraction begin to degrade the detectability inside approximately an arcsecond.  
However, for 4.8 \micron\, observations, the brighter sky background  and more stable PSF at longer wavelengths
result in the images being sky-background limited even for the closest
planet separation.

\section{Data Reduction and Analysis}

The set of data was reduced with a custom-developed pipeline using the
Perl Data Language (PDL\footnote{http://pdl.perl.org}).  
A cube of frames for each observation is coadded to obtain individual 
images of 15.3, 14.6, or 24.0 s for the 3.3 \micron, 
\lprime\, and M band observations respectively. Each
image is subtracted from its pair in the other nod position to create
two sky-subtracted images of the star.  To eliminate the effect of
slightly different gains among the four different channels, a blank
portion of the detector is used to measure the average flux in each
channel. This value is subtracted from all pixels corresponding to
the appropriate channel.  Finally, a bad pixel rejection routine is
carried out which looks at the detector response for two successive
nods to identify pixels which vary by more than 3 times the standard
deviation.  The values for these pixels are replaced by an average of
their neighbors.  The images are then shifted by measuring the
centroid of the star and rotated by the Parallactic angle of the
observations to create a final image of the star with North up and
East to the left in the image.  A median PSF image is created by
median combining all of the images without rotation.  By subtracting this
median PSF from all of the images prior to rotation, the majority of
the diffraction and static aberration-induced scattered light can be
removed from the final, combined image.

The measurements of the brightnesses of the planets are calculated relative to
the measured flux from HR 8799 itself.  Since the star is saturated in
the \lprime\, and 3.3\,\micron\, images, images taken with neutral density
filters or shorter exposures were obtained to measure the flux.  The photometry 
is scaled from the 2MASS K$_s$ band flux of HR 8799, using the expected K-L' and 
K-M colors for HR 8799's spectral type \citep{Marois08, Cox00}, see Table \ref{par}.

The scale of the Clio pixels has been measured to be 0.04857\,$\pm$\,0.00003 "/pixels, 
using periodic observations of binary stars \citep{Heinze07}.  The orientation of the detector 
relative to the sky is also calibrated with these measurements. North on the detector is 
92.27$\pm$0.2 degrees counter  clockwise from the top of the array.

\section{Observational Results}

To calculate the flux of the planets, apertures of sizes of 0.25
arcsec were placed at the approximate locations of the sources.  The
images of the stars were measured with the same aperture to account
for aperture-dependent effects.  To estimate the errors in these
measurements, each data set was split in half.  For each half of the
data set a measurement of the image in each nod position was carried
out, resulting in four independent measurements.  The apparent magnitude and standard error
in the mean for each set are reported in Table \ref{phot}.  For M band, and some of the
3.3\,\micron\, data the planets are not detected.  For these data, 3
$\sigma$ upper limits are derived by measuring variations in flux with the same size aperture, 
at the same radial separation as the planet in an azimuthal arc around the star.   We note that, typically, we conservatively
quote 5-7 $\sigma$ limits for detection with Clio observations. However, for these
observations, where we know the exact location of the planet, 3 $\sigma$ limits are more
realistic estimates of our uncertainty in the flux contribution.

For comparison with models and other photometry we adopt a distance of 39.4 pc to HR 8799
\citep{vanLeeuwen07}.  The photometric measurements of planets $b$, $c$, and $d$ are reported 
in Table \ref{phot}.  We tabulate absolute and apparent magnitudes, as 
well as the "absolute" spectral flux density in milliJanskys, or the
flux at a distance of 10 pc.  This is useful for comparisons with model spectral energy
distributions. The \lprime\, band measurements are consistent with \citet{Marois08}, within our photometric
uncertainty.

The initial observations at [3.3] detected $c$, but did not detect the other two components.  Follow-up
observations in September 2009 confirmed the detection of $c$, and also detected $d$. We report the photometry
for 3.3 \micron\,from the September 2009, as it is more sensitive than the January 2009 data, with better PSF subtraction.
For the September 2009 data, there is a marginal source at the location of $b$. However, it is below the significance level
of the image. We thus adopt the limiting magnitude for the image as the upper limit for $b$ at 3.3 \micron.

The astrometric measurements are calculated with a "center-of-light"
algorithm within the photometric aperture.  The same procedure of
dividing up the data into four independent sets is used to calculate
the astrometric uncertainty.  These uncertainties are consistent with
the astrometric measurement errors expected due to the PSF width
of these observations and their low signal-to-noise ratio
\citep{Lindegren78}.  Astrometric results for the observations are summarized in
Table \ref{ast}.  The \lprime\,position of planets $b$ and $c$ are consistent
with the Marois results, differing by 24 and 44 mas respectively.
The \lprime\,  $d$ position is displaced by 62 mas from the reported Marois
position. Since the orbital motion expected for this object is only 7
mas over the two month time span between the two observations, the
 discrepancy of $d$ is not easy to explain.  It is possible that the 
 positional difference is an indication of contamination of the
flux by a residual speckle in the final image.   The [3.3] position of $d$ is
also 66 mas from the Marois position, but since the observations were taken
a year after those results, the difference is similar to the expected orbital motion.
This effect can be seen in Figure \ref{imfig}, where the position of $d$ in the [3.3] image
is significantly shifted relative the the \lprime\, image.

\section{On the Age and Stellar Multiplicity of HR 8799}

Understanding the age of the HR 8799 system is important for interpreting the SEDs
of the planetary companions and assessing their stability.  For example \citet{Fabrycky09} and \citet{Reidemeister09} 
analyze the stability of this system, for which one of the crucial parameters its age.  \citet{Marois08} 
assign an age for HR 8799 of 60$^{+100}_{-30}$ (30-160) Myr, and previously \citet{Moor06}
estimated an age of 20-150 Myr based on assigning HR 8799 membership
to the Local Association. \citet{Marois08} based the age on careful
consideration of four diagnostics:  1) the star's space motion and possible association 
with HD 984 and HD 21318, as well as the Columba and
Carina Associations, 2) its position in a color-magnitude 
diagram, 3) the typical age of $\lambda$ Boo and $\gamma$ Dor variables, and 4) the large mass of its debris
disk.  Of these, only the star's position on a color-magnitude diagram provides
a quantitative constraint.  Here, we explore whether 
association of HR 8799 with any additional companions or moving 
groups can be used to further refine its age estimate. 

\subsection{On the Association with HD 984 and HD 21318}

\citet{Marois08} calculate a space motion for HR 8799  of UVW = (-11.9, -21.0, -6.8)
km/s, and suggest that it is similar to other young stellar
groups in the solar neighborhood (e.g.  Columba, Carina, etc.)  as
well as two neighboring stars: HD 984 and HD 21318 to which they assign
ages of $\sim$30 and $\sim$100 Myr, respectively. We investigate the space motion of HR8799
here and discuss Columba and Carina groups in the following subsection.
 Using the revised Hipparcos astrometry
\citet{vanLeeuwen07} and the radial velocity from the compiled RV
catalog of \citet{Gontcharov06}, we estimate the velocity of HR 8799
to be UVW = (-12.2, -21.2, -7.2) $\pm$\, (0.6, 0.8, 1.2) km/s. Using
astrometry from \citet{vanLeeuwen07} and radial velocities 
from \citet{Holmberg07}, we independently verify
that the stars HD 984 and HD 21318 have space motions 
within 3.6\,$\pm$\,2.0 \kms\, and 1.0\,$\pm$\,1.8 \kms\, of HR 8799,
respectively. The HD 984 and HD 21318 are separated by 26 and 19
pc from HR 8799, respectively. We simulated the orbits of HR 8799 and
these two stars using the epicyclic approximation, adopting
the Oort constants from \citet{Feast97} and the LSR velocity from
\citet{Dehnen98}, and the local disk density from
\citet{vanLeeuwen07}. We find that HD 8799 and HD 984 were not significantly
closer to one another within the past 100 Myr (although some
$\sim$0.02 pc closer $\sim$0.5 Myr ago), and were probably separated
by $\sim$80 pc $\sim$30 Myr ago. We find that HR 8799 and HD 21318
would have been separated by $\sim$140 pc $\sim$100 Myr 
ago, with the nearest pass being $\sim$13 pc away $\sim$11 Myr ago. While
$\sim$1 km/s uncertainties in the velocities can lead to large
uncertainties in the distant past ($\sim$1 pc/Myr), any genetic tie between HR 8799 and 
HD 984 and HD 21318 would seem tenuous at best. 

\subsection{On the Association with Columba and Carina}

Is HR 8799 related to the putative Columba or Carina groups?  
We adopt our revised space motion for HR 8799 and 
the parameters of the velocity and position centroid for Columba
and Carina from \citet{Torres08} and ran the stars through 
our epicyclic orbit code. HR 8799 is currently 98 pc from the
centroid of Columba, as defined by \citet{Torres08}, and its 
closest approach to the group within the past 160 Myr was a pass $\sim$27 Myr
ago ($\Delta$ $\simeq$ 58 pc), which seems too distant to be a
plausible candidate for ejection/evaporation. Similarly, HR 8799 is
currently $\sim$116 pc distant from the centroid for Carina, and its
closest approach within the past 160 Myr was $\sim$37 Myr ago ($\Delta$
$\simeq$ 45 pc) -- also relatively distant for a plausible association. We
find that the centroids of Columba and Carina were closest to one
another $\sim$18 Myr ago with a separation of $\sim$24 pc. While the
putative groups themselves are very large (tens to $>$100 pc), it 
is more likely that they constitute ``complexes'' with a wide range of
ages, making membership -- even if likely -- not useful for assigning ages.

\subsection{On the Candidate Common Proper Motion Stars Found by
Close \& Males (2009)}

\citet{Close09} found no evidence for bound companions using archival
Gemini and HST/NICMOS images. They also identified a few candidate
common proper motion stars within a degree using NOMAD/USNO-B1.0
astrometry \citep{Monet03, Zacharias05}, however none are particular
strong candidates for being associated with HR 8799. Despite the
similarity of their proper motions to that of HR 8799, the two top
co-moving companion candidates identified by \citet{Close09} can be ruled out through 
further consideration of their photometry. NOMAD \#1115-0634383 (J2000; 23 10 22.93 +21 34 20.7) is
48' distant from HR 8799 and is very red. We find that the 2MASS JHK
photometry for this object (J-H = 0.58\,$\pm$\,0.04 mag, H-K$_s$ =
0.22\,$\pm$\,0.04 mag) is consistent with an unreddened M2V ($\pm$1
subtype). This corresponds well with the B-V estimated by
\citet{Close09} (B-V $\simeq$ 1.61; inferred from NOMAD photographic
magnitudes) and with a V-K$_s$ color ($\simeq$ 4.5 mag) that we
calculate using USNO-A2.0 B \& R photometry
\footnote{http://www.projectpluto.com/photomet.html} and 2MASS K$_s$.
However, if the star is an unreddened M1V-M3V dwarf at d = 39 pc
(M$_V$ $\simeq$ 13.9 mag), it is approximately 2-4 magnitudes below
the main sequence, and hence could not be related to HR
8799. Similarly, the second best candidate companion NOMAD
\#1108-0634609 (TYC 1717-1120-1) can be rejected on similar
grounds. The V-K$_s$ color \citep[1.87\,$\pm$\,0.04; ][]{Cutri03,
Droege07} is consistent with an unreddened G8V ($\pm$1 subtype).  At a
distance of 39 pc, the observed V magnitude \citep[V = 10.50
;][]{Droege07} would correspond to M$_V$ = 7.52, which is $\sim$2
magnitudes fainter than typical G8V dwarfs. The colors of both
candidates are also far too red to correspond to unreddened young
($\sim$10$^{7-8}$ years) degenerate stars \citep[see
e.g.][]{Kalirai03}. Reddening is unlikely to be responsible for the
brightness deficits of these two objects: reddening is negligible
within 75 pc of the Sun \citep[e.g.][]{Leroy93}, and HR 8799 itself
has negligible reddening \citep{Zerbi99}. We conclude that both of the
two strongest NOMAD candidates identified by \citet{Close09} are probably 
background field stars, unassociated with HR 8799.

\subsection{Search For Co-Moving Stars Within Ten Degree Radius}

Some of the famous young nearby stars in the solar vicinity have been 
found to belong to stellar associations which help in their age-dating
(e.g. TW Hya, AB Dor), mostly through combining X-ray and astrometry
data to search for active, co-moving stars \citep{Zuckerman04}. This
motivated us to search for evidence of a group in the vicinity of HR
8799 similar to other studies above. At present, we focus on X-ray/astrometric selection since
astrometric selection alone is rather daunting and produces a substantial chance of false positives \citep[e.g.][]{Close09}.  We
searched a 10$^{\circ}$ ($\sim$10 pc) radius around HR 8799 for signs
of co-moving low-mass stars with a cross-referenced list of
X-ray-emitting stars \citep{Voges99, Voges00} in the Tycho-2 and UCAC2 astrometric catalogs
\citep{Hog00, Zacharias03}. As the 3D velocity of HR 8799 is well-defined,
we tested whether the motions of these stars were consistent with
motion towards HR 8799's convergent point ($\alpha$, $\delta$ =
99$^{\circ}$.5 $\pm$ 2$^{\circ}$.8, -31$^{\circ}$.1 $\pm$\,
1$^{\circ}$.8 deg). The most interesting system identified was the dMe
star [K92] 94 (= 2MASS J23081213+2126198)\footnote{The originally
published position from \citet{Kun92} is apparently off by
$\sim$30''. We correctly matched the 2MASS source to the star using
the finder chart in Fig. 8 of \citet{Kun98} and SkyView
(http://skyview.gsfc.nasa.gov/).} situated 1235'' away from HR
8799. The star was originally discovered due to modest H$\alpha$
emission by \citet{Kun92}, and later classified as M0V without
H$\alpha$ emission by \citet{Martin96}. However, if placed at 39 pc,
its absolute K magnitude (M$_K$ $\simeq$ 8.1; based on 2MASS K$_s$
photometry) would be $\sim$3 magnitudes fainter than a typical young
disk M0V star \citep[M$_K$ $\simeq$ 4.8;][]{Leggett92}.  Combined 
with the fact that its tangential motion differs from that of HR 8799 by
$\sim$4 km/s, it seems unlikely that [K92] 94 is associated with HR 8799. 

We therefore conclude that there still is no evidence for any co-moving stellar companions to 
HR 8799 which could help constrain its age. For the purpose of estimating planet 
masses we thus use the same age estimate as \citet{Marois08} of 60$^{+100}_{-30}$ Myr.

\section{Comparison to Giant Planet Models and Brown Dwarfs}

The observational results above can be used to constrain the physical parameters
for the planets around HR 8799.  We have used two separate approaches:
comparison with young, giant planet models in the literature, and comparison
with field brown dwarf photometry and spectra.  

For the giant planet models, we can use the evolutionary cooling tracks of \citet{Burrows97}, hereafter B97, 
to estimate the size of the planets.  The modeled radii of objects in the mass range of 5-20 \mjup\, 
and an age range of 30-160 Myr vary from 1.15-1.29 \rjup.  We adopt a radius of 1.22 \rjup to estimate
the flux density from the model SEDs.  This size 
suggests a gravity of \logg=4.0-4.6 is expected for these objects.  We use models of \logg=4.5 for comparison 
with the observations.

Because of the small variations in radius versus age and mass, the key observable to constrain is the effective
temperature of the objects.    Depending on the age of the system, the effective temperatures 
will correspond to different mass estimates.

\subsection{Giant Planet Models}

Models of brown dwarf and giant planet atmospheres have been carried
out with increasing levels of detail over the past decade.  Non-gray atmospheres
have been computed by \citet{Burrows97, Burrows03}, and by \citet{Baraffe03} 
that cover this temperature range.  The effects of both clouds and non-equilibrium
chemistry have been studied by \citet{Saumon06} and \citet{Hubeny07} .  
\citet{Hubeny07} have developed a grid of models using various
diffusion coefficients, reaction rates for CO $\rightarrow$ CH$_4$ and
gravities.  They find that the M band flux, in particular, is affected
for \teff\, $\sim$ 1000 K, exactly the range for the three HR 8799
planets.   \citet[][and references therein]{Saumon08} have developed models 
that incorporate vertical mixing \citep{Saumon06, Leggett07}  and increased metallicity 
\citep{Fortney08} to match spectra of brown dwarfs and predict the SED's of giant planets.  
We compare the observations above to families of models by \citet{Saumon06} (hereafter S06)
and \citet{Hubeny07} (hereafter HB07) that include these effects.

\begin{figure}
\epsscale{.95}
\plotone{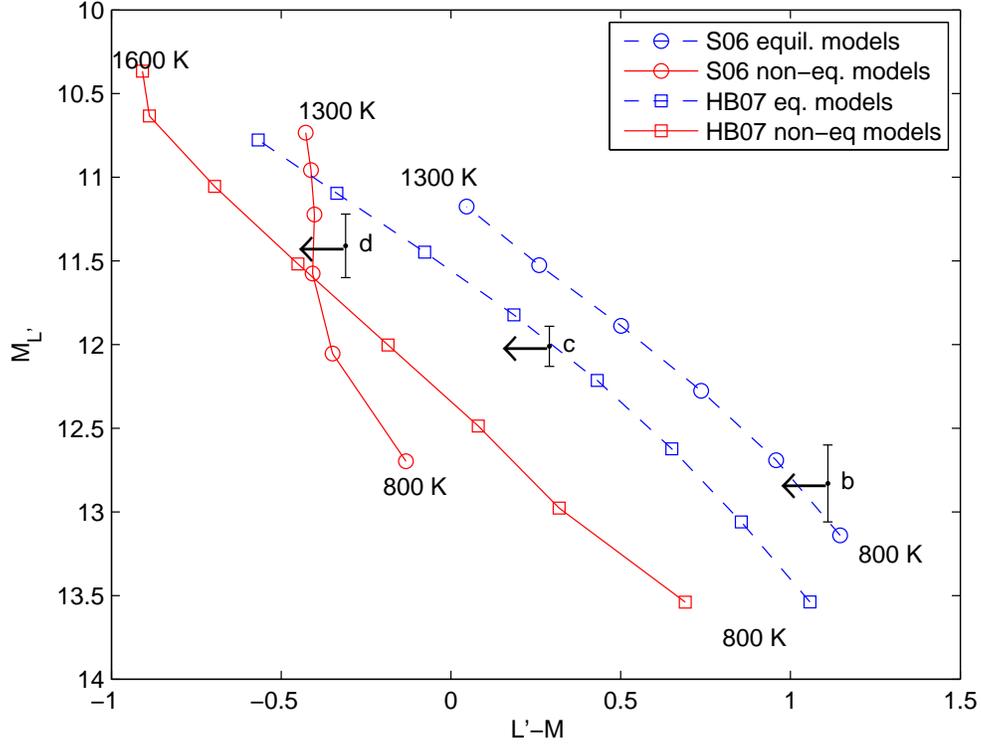}
\caption{Color magnitude relation for models compared to observed photometry. Model colors are 
shown from \citet{Hubeny07} and \citet{Saumon06}.  The temperature range spans 800-1600 K for the
HB07 models and 800-1300 K for the S06 models.  The models use K$_{zz}$=10$^4$ cm$^2$ s$^{-1}$,
 \logg=4.5, and a cloudy atmosphere model.  A chemical reaction rate corresponding to the "slow" rate
 is used for the HB07 models, and the "fast2" rate is used for the S06 models.
\label{colmag}}
\end{figure}

We plot a color vs. magnitude at \lprime\,  and M in Figure \ref{colmag}  to compare
the observations with the models.    The M band limits are the three $\sigma$
limits listed in Table \ref{phot}.   The data are compared to both models that are in
chemical equilibrium and models which incorporate non-equilibrium chemistry.
While $b$ could be fit by any of the model sets, the location of $c$ is inconsistent
with the equilibrium S06 models, and $d$ is inconsistent with both sets of
equilibrium models.   For both the HB07 models and the S06 models, the non-equilibrium
 sequence uses an eddy diffusion coefficient  (K$_{zz}$) of 10$^4$ cm$^2$ s$^{-1}$.  The HB07 models shown use the
 "slow" reaction for the CO $\rightarrow$ CH$_4$ chemical reaction described in \citet{Hubeny07}.  The S06 models use 
 an equivalent rate to the "fast2" rate described in HB07 (D. Saumon, private communication).  ,  The models incorporate clouds 
 and are calculated for  \logg=4.5   The temperatures for
each sequence are in 100 K increments, from 800 K to 1600 K for the HB07 models and 800 to 1300 for
the S06 models.  The colors suggest that, as with brown dwarfs \citep{Leggett07}, the HR 8799
objects are best fit by models which incorporate vertical mixing into the atmospheres.

\begin{figure}
\includegraphics[angle=0,scale=.90]{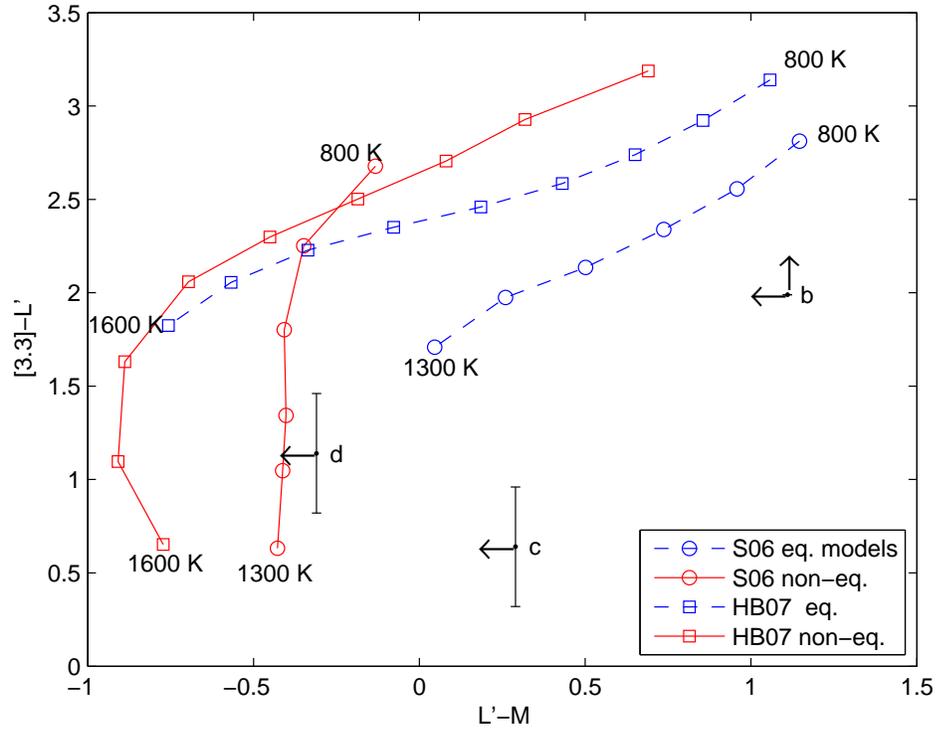}

\caption{Color-Color plot for \lprime-M vs. [3.3]-\lprime.  Models from \citet{Hubeny07}
and \citet{Saumon06} are shown. The same parameters are used as in Figure 2.  
\label{colcol}}
\end{figure}

Figure \ref{colcol} shows  a color-color plot for the object and models, using   \lprime-M vs. [3.3]-L.  The models shown are
the same as for Figure \ref{colmag}.   Due to its faintness,
the limits on $b$ are consistent with any of the models.  The [3.3]-\lprime\, color of $c$ and $d$
is bluer than predicted by the equilibrium models over the range of temperatures considered.
The HB07 non-equilibrium models  are only consistent with the [3.3]-\lprime\, color for
effective temperatures above 1500 K, for $c$, and 1430 K, for $d$.  The S06 non-equilibrium models are well
matched to colors,   for $c$, which correspond to  1200-1350 K and, for $d$, 1050-1250 K. For $b$, the effective temperature 
estimated from the S06 model colors is below 950 K.  In general, the S06 models indicate cooler temperatures by $\sim$300 K
for all three planets, compared to the HB07 models.

\begin{figure}
\epsscale{.95}
\plottwo{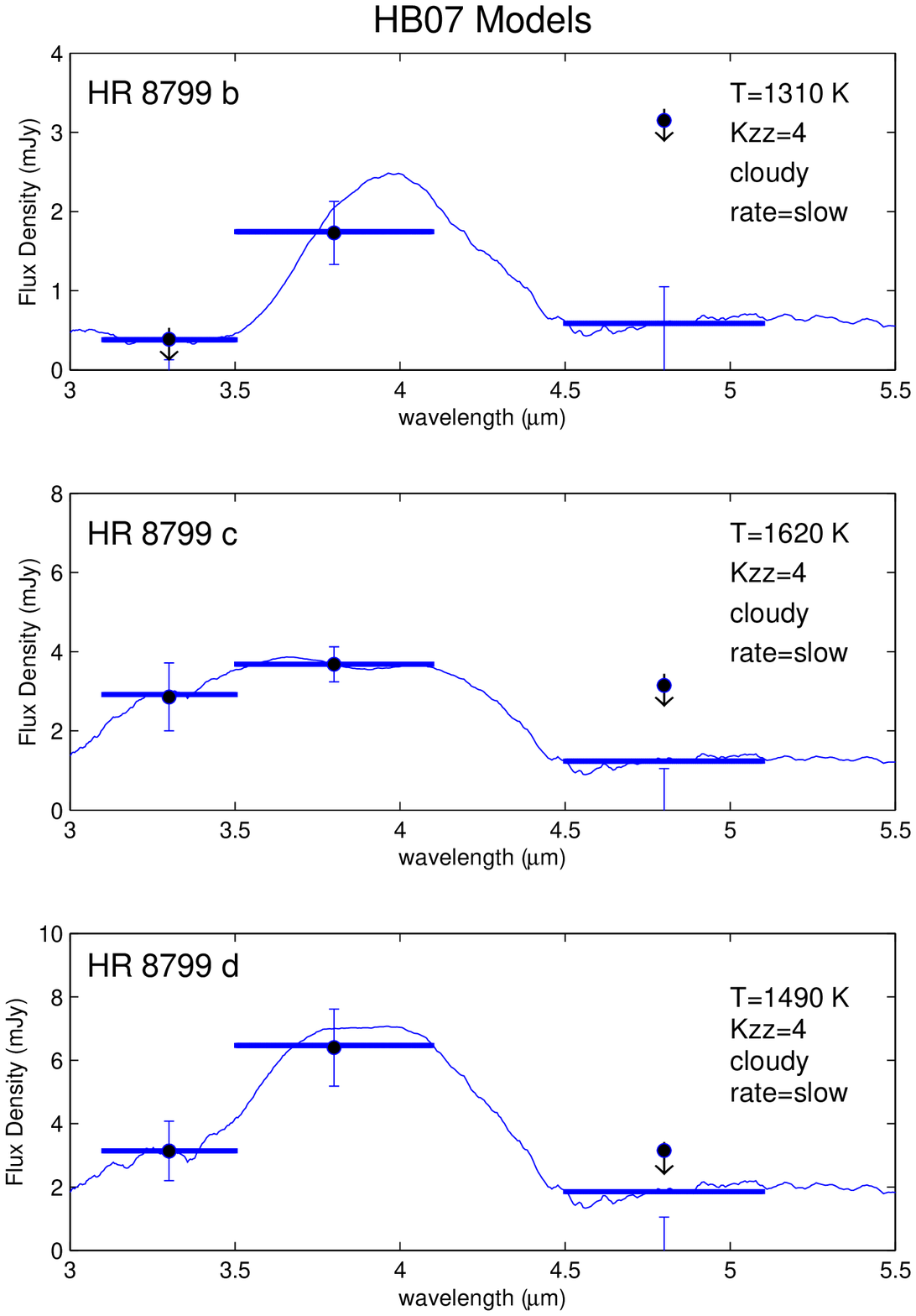}{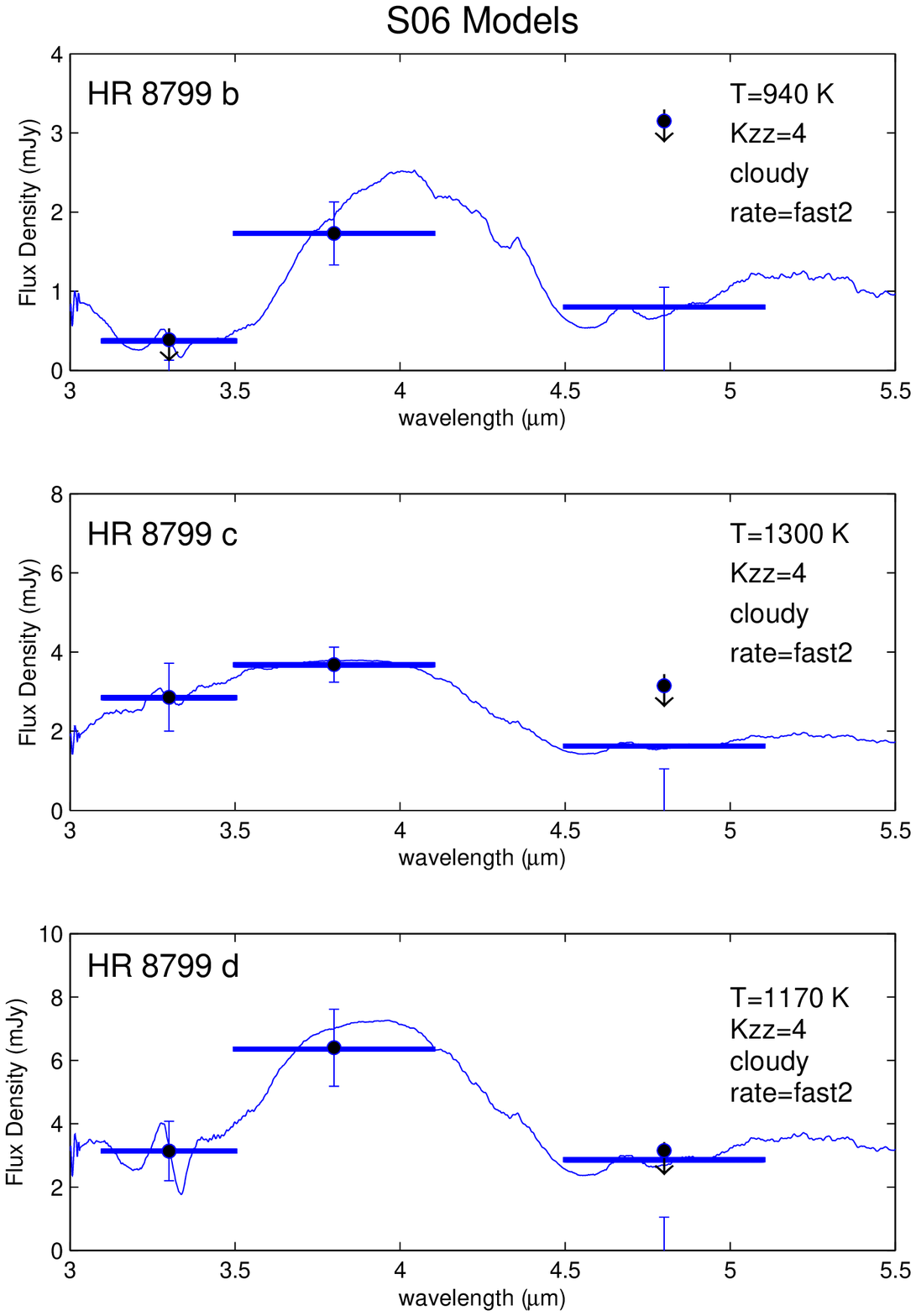}
\caption{Comparison of HR 8799 photometry to models by \citet{Hubeny07} and \citet{Saumon06}
for $b$(top), $c$(middle), and $d$(bottom).  For each planet, models are
shown that best match the [3.3]-\lprime\, color.  The overall flux is allowed to vary to fit both the [3.3] and \lprime
photometric points. The band-integrated flux  density of the models are shown as horizontal lines for the 
observed passbands.  
\label{modelcols}}
\end{figure}

Model fits to the [3.3]-\lprime\, color can be used to estimate the effective temperatures of the planets.  These fits are shown in Figure 
\ref{modelcols}.  The HB07 models result in \teff $<$ 1310 K for $b$, and \teff=1620 and 1490 K for $c$ and $d$. For the S06 models, 
the corresponding results are $<$940 K for $b$, and 1300, and 1170 K for  $c$ and $d$, respectively.   

The model fits to the colors do not constrain the overall flux. If a size of 1.2 \rjup\, is used to estimate the flux, consistent with
expected size of these objects from the B97 models, the results are significantly brighter than the observed photometry at \lprime\,.
 To match the observed flux at the best fit temperatures, the planets would need to be smaller than expected from the B97 models.  
 For the HB07 models, planets $c$ and $d$ would need to have radii which are 44 and 63\% of 
their expected size of 1.2 \rjup.  For the S06 models the radii would need to be 56 and 84\% as big.  

\begin{figure}
\epsscale{.95}
\plottwo{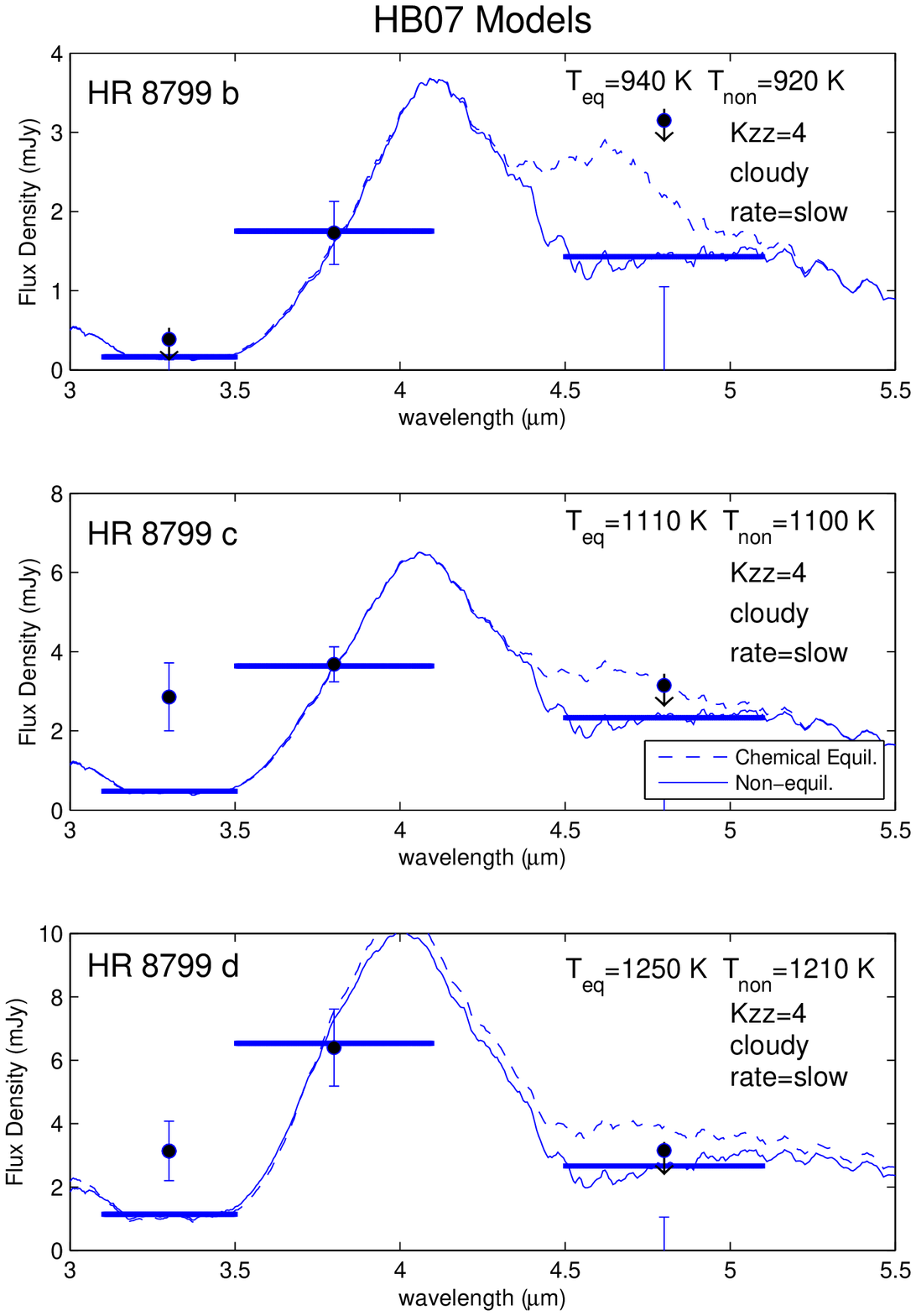}{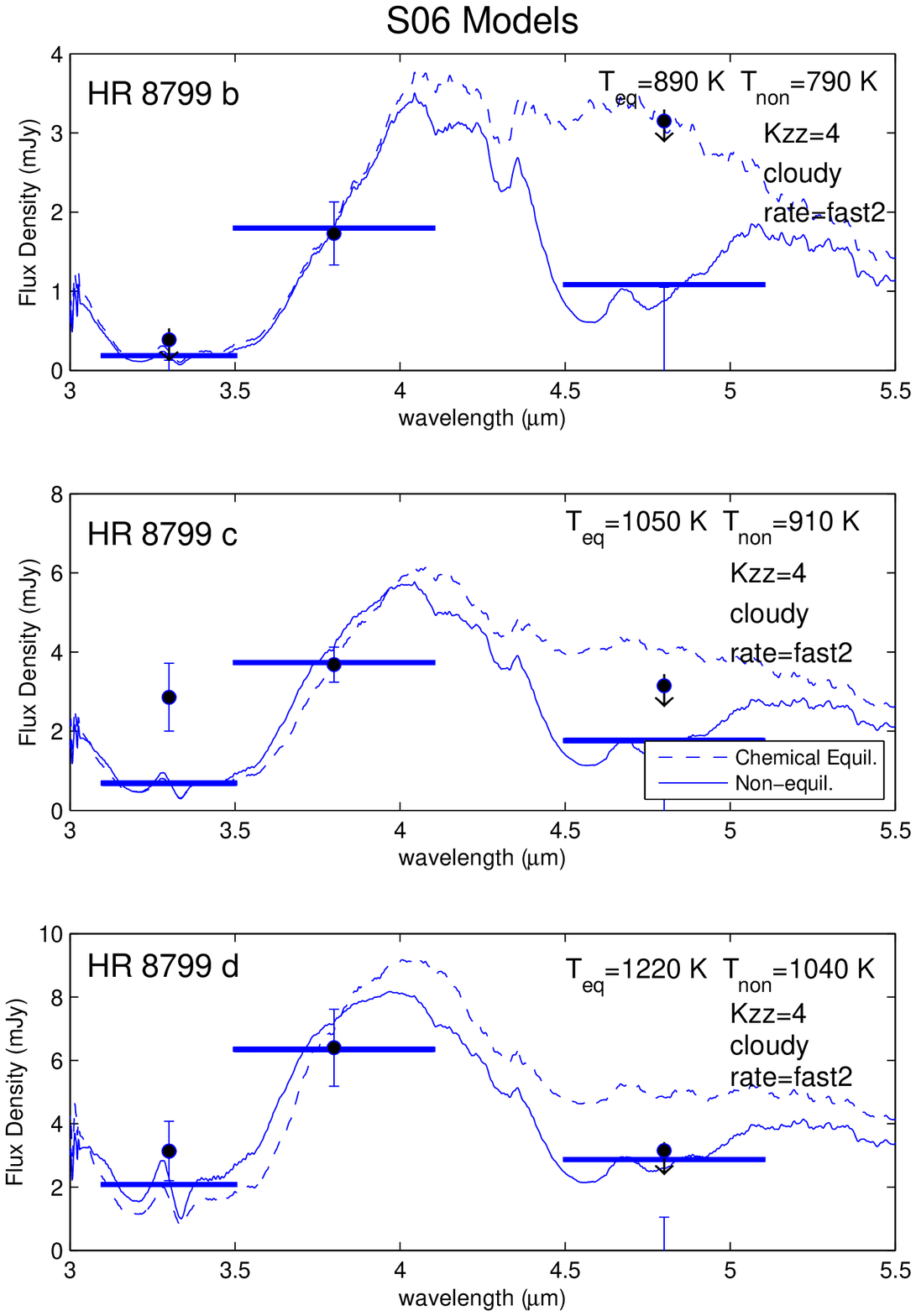}
\caption{Comparison of HR 8799 photometry to models by \citet{Hubeny07} and \citet{Saumon06}
for $b$(top), $c$(middle), and $d$(bottom).  For each planet, models are
shown that best match the \lprime\, band photometry.    
The chemical equilibrium models appear to over-predict the M band flux, while
non-equilibrium models are consistent with the 3 $\sigma$ upper limits.  The band-integrated 
flux  density of the non-equilibrium models are shown as horizontal lines for the 
observed passbands.  
\label{models}}
\end{figure}

Figure \ref{models} again shows the SEDs of the HB07 and S06 models compared to the HR 8799
photometry, but now matched to the absolute magnitude at \lprime\,(\mlprime).  For each planet, a model temperature is selected that best 
fits the value of \mlprime, using an assumed radius of 1.22 \rjup\,  for the planets.    The effective temperature and model parameters corresponding 
to the fit for each plot is inset in the graph. For the non-equilibrium SEDs, the average modeled flux in each passband is plotted  as a horizontal line spanning the
passband.  The difference in the 
best fit SEDs between the HB07 and S06 models suggests that the details of how the vertical mixing is modeled is important 
for accurate estimates of colors, especially in the 3.3 \micron\, portion of the spectrum.  

\begin{figure}
\includegraphics[angle=0,scale=.50]{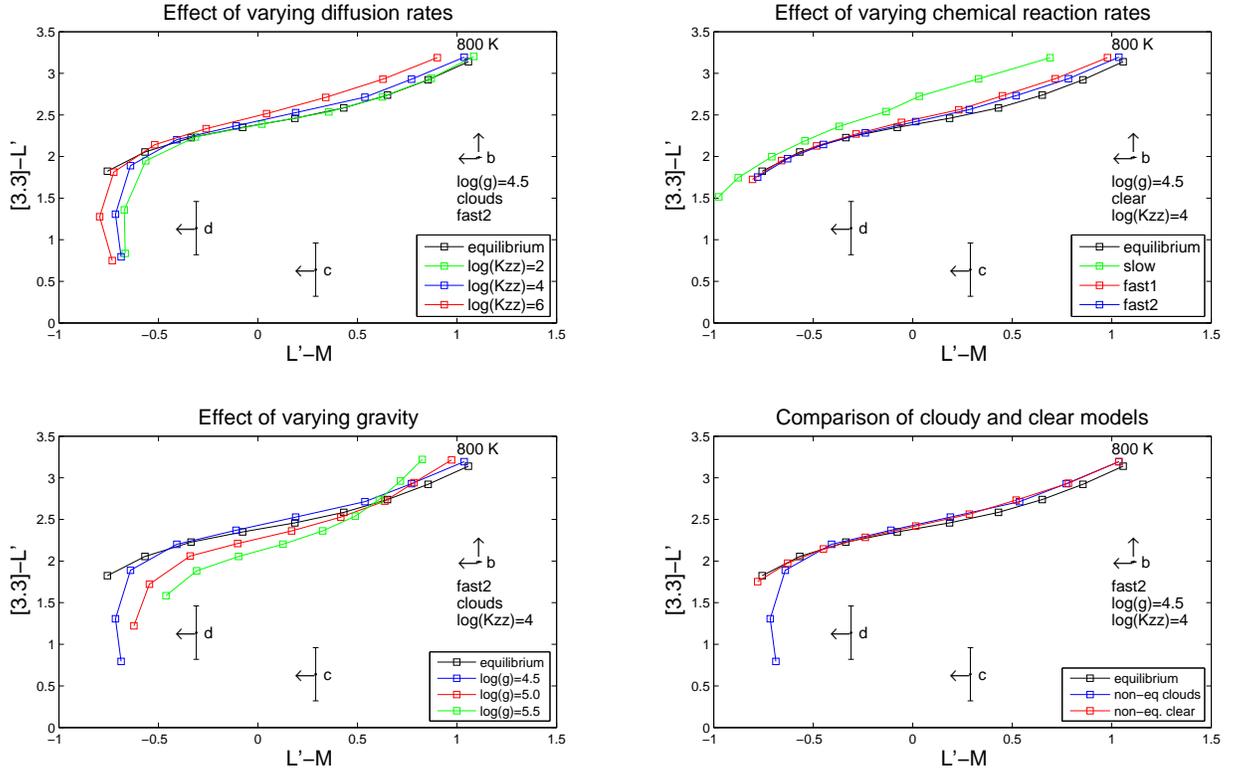}

\caption{Color-Color plots using the \citet{Hubeny07} models.  The upper left compares different diffusion rates.  The right
compares different reaction rates.  The lower left compares varying gravity, while the lower right compares cloudy and clear
models.  The models are unable to reproduce the [3.3]-\lprime colors unless temperatues above \teff=1400 K are used.  
\label{colcolcheck}}
\end{figure}

Different diffusion rates, reaction rates and gravities ,as well as a comparison of cloudy and clear models were explored
for the HB07 models.   These comparisons are shown  in Figure \ref{colcolcheck}.  Only cloudy models are available for the various
reaction rates tried, while the varying diffusion rates are only calculated for clear models.  While this leads to an imperfect comparison, the general
trend for each parameter can be seen.   Cloudy models can reproduce the [3.3]-\lprime, colors seen for $c$ and $d$, but only for \teff$>$1400 K.   
The other parameters do not significantly affect the [3.3]-\lprime\, colors.  

\subsection{Comparison to Brown Dwarf Spectra}

Another useful comparison to the observational results can be found in the
 colors of brown dwarfs in the field.   Although field brown dwarfs of similar temperature 
 to the HR 8799 are likely more massive, due to being older, they might still be expected to provide a
good approximation of the expected SEDs of these objects.

Although the  [3.3]-\lprime\, colors are blue relative to most model SEDs, similar colors
are seen in spectra for brown dwarfs at these effective temperatures.  In addition, the observed [3.3]-\lprime\,
color changes significantly for spectral types in the L9 to T3 range, confirming that this region of the spectrum
is useful for constraining the physical parameters of brown dwarf and giant planet atmospheres in this temperature range.
For example, \citet{Leggett08} analyze HN Peg B in detail and \citet{Stephens09} present infrared 
spectra of a large sample of brown dwarfs, including, 3-4 \micron\,spectra of a subset of 
the objects.  The objects SDSS 1207+02, SDSS 0758+32, HN Peg B, and 2M 0559-14 with spectral types of T0, T2,
T3, and T4.5 have [3.3]-\lprime\, colors of approximately 0.75, 1.2, 1.9, and 2.5 respectively.  
HR 8799 $c$  and $d$ have a [3.3]-\lprime\, color of 0.63 and 1.16 respectively.  
This suggests that the [3.3]-\lprime\, color of  $c$ and $d$ are consistent with a T0$\pm$1 and T2$\pm$1 
dwarf, respectively.  From the temperature scale of  \citet{Golimowski04} as revised by
\citet{Stephens09} this would indicate \teff=1200-1300 K,  for these objects, 
consistent with the S06 models.    $b$ has a [3.3]-\lprime\, color of  $>$ 1.9 
constraining its spectral type to be later than T2, or \teff$<$ 1200 K. Of course, the \citet{Stephens09}
 \teff\, vs. spectral type relation shows scatter of $\sim$100 K about a given spectral type, 
 suggesting caution should be used in comparing these results,

Finally, we note that the H-[4.5] color-temperature relation
developed by \citet{Warren07} can be used to constrain the temperature
of these objects, assuming they have similar emergent spectra to field
brown dwarfs.  \citet{Leggett07} derive a constant [4.5]-M'=-0.3 color correction
for objects in this temperature range.  From the \citet{Marois08} H band photometry
we calculate an H-[4.5] color of less than 2.47, 1.63, and 1.56 for $b$, $c$, 
and $d$ respectively.  Using the Warren scale we derive temperature greater than 900, 1100, and 1150 K
for the objects, consistent with the color-derived estimates from the S06 models.




\section{Discussion}

The key feature of the 3-5 \micron\, results is the blue color of the
photometry, compared to equilibrium models.
While the \lprime-M colors can be matched with the HB07 models, the [3.3]-\lprime\, model colors are
too red unless effective temperatures of $>$1400 K are used.  
Vertical mixing can explain both the M band and 3.3 \micron\,observations
in the S06 models.  The difference between these similar model families suggests that the details of how 
the non-equilibrium chemistry is modeled appears to be an important factor in the resulting SED.  \citet{Janson10} reach a similar 
conclusion from their analysis of $c$ s spectrum, in the L band.  



Even the S06 models  have a discrepancy between the effective temperature derived from the
[3.3]-\lprime\, color and the best fit according to the \lprime\, photometry, with the \teff\, predicted to be
$\sim$ 100-400 K lower from the \lprime photometry compared to the color-derived temperature. 
The correspondence of the color-derived temperatures from S06 models with those of field brown dwarfs
suggest that the model temperatures are reasonable.  Based on the effective temperatures, the derived
masses from  the cooling models of \citet{Baraffe03} are 12$\pm$2 \mjup\, for $c$ and 11$\pm$2 \mjup\, for $d$, 
for an assumed age of 60 Myr.  The temperature limit on $b$ results in a mass limit of $<$ 9 \mjup.    
  These mass estimates are consistent with the estimates from \citet{Marois08}.  

 The large age range for the HR 8799 system (30-160 Myr) increases the uncertainty in the estimated mass of these objects. This error term
 is larger than that from the photometry, resulting in a mass range of $\sim\pm$3 \mjup\, for the objects.  Combined with the photometric uncertainty, this
 indicates a mass of 12$\pm4$ \mjup\, for $c$, 11$\pm$4 for $d$, and an upper limit on $b$ of $<$ 12 \mjup.

The dimness of $c$ and $d$, given the color-derived temperatures is problematic for any evolutionary model of
a giant planet. Even a very old object will not be small enough to match the \lprime\, apparent magnitudes. The best fits
to \teff\, of $c$ and $d$ result in radii for the objects of 0.7 \rjup\, and 1.0 respectively.  Given the young age of
HR8799, a possible explanation of this discrepancy is that dust extinction of approximately 1.3 and 0.4 magnitudes
at \lprime\, is reducing the observed flux for $c$ and $d$ respectively.  Since HR 8799 is relatively nearby an shows no sign
of foreground extinction, this dust obscuration would need to be intrinsic to the objects themselves.

Alternatively, lower gravity, metallicity, details of the chemical abundance versus height, or even incorrect opacity estimates for methane may be affecting
the colors.  Models with \logg=4 for the S06 were also compared to the observations, but were not different from the \logg=4.5
models shown.  However, it is possible that gravity, combined with effects not properly captured in these models (such as metallicity
or details of the chemical abundance variation versus height) may be affecting the colors for these objects in a more significant way than for field
brown dwarfs.    Typical field brown dwarfs have median ages of $\sim$3 Gyr
\citep{Dahn02}. Hence, the typical gravity of a \teff\,= 1000\,K object in
the field is \logg\, = 5.0, while these objects are expected to have  \logg\,= 4.3.  If the vertical mixing is similar for both the
field brown dwarfs and the planets, it is reasonable to expect that
the stronger than expected CO absorption, and the weaker than expected
CH$_4$ absorption is an effect of the lower gravity.

Interestingly, the NIR colors of these objects, as well as 2M1207b, are redder than field
brown dwarfs of the same brightness \citep{Marois08}. The NIR color difference
could well be explained by their youth, and thus their lower gravity. Indeed,
the young field L dwarf 2M0141-4633, found by \citep{Kirkpatrick06}, has nearly
identical NIR photometry to $c$ and $d$,and is discrepant with other field L dwarfs.
The estimated mass of 6-25 \mjup for this object indicates it should have similar 
gravity to the HR 8799 planets.  This suggests that the red NIR colors are a low gravity
effect. Kirkpatrick et al. attribute the red colors to lower collision-induced absorption of H$_2$, 
which primarily suppresses the K band. At lower gravities the H$_2$ CIA is reduced, causing 
the redder H-K color.  Thus, while the color trend is opposite in the thermal IR, it appears that both
effects might be explained by the youth, and low gravity, of the objects.   

While in principle, the balance between CO and CH$_4$ in these atmospheres is governed by
Le Chatelier's principle \citep{Burrows01},  the details of how the vertical mixing occurs, appears
to be important in predicting the emergent spectra.  An additional effect may be the metallicity of the
atmospheres \citep{Fortney08}. In higher metallicity objects, the formation of CO might expected to be 
enhanced since more oxygen is available relative to hydrogen, compared to lower metallicity atmospheres.
This suggests that giant planet atmospheres would be blue in the 3-5 \micron\, region for more metal-rich objects.

Although HR 8799 is metal-poor star, detailed observations \citep{Gray99} indicate solar abundance for lighter 
elements \citep{Sadakane06}, a consequence of the $\lambda$ Boo phenomenon.  This suggests that models
with solar metallicity are probably appropriate for interpreting the colors of the HR 8799 planets.  The solar
metallicity (or perhaps metal-poor if the $\lambda$ Boo phenomenon is discounted) cannot explain the blue 
3-5 \micron\, colors.  If the formation of the planets caused them to be significantly metal-enhanced relative to their
stars, this may be able to explain their colors.  Further observations may be able to disentangle the metallicity
of these objects from the other parameters which govern their SEDs.   At the same time, detailed models are needed 
to explore the  potentially complex interplay between gravity, vertical mixing, and metallicity.

It is interesting to note that mass estimates, derived from model fits to the \lprime\, photometry are not significantly different from
color-derived estimates.
 For example, the non-equilibrium fits from the HB07 models, for an age of 60 Myr, result in
mass estimates of 9, 10.5, and 11 \mjup\, for $b$, $c$, and $d$, respectively.  If we use the S06 models, the estimates are 7, 9, and 10
\mjup  respectively.     The HB07 are likely a better estimate of the effective temperature 
from the \mlprime values, since these models treat the temperature profile
in a self-consistent way with the vertical mixing, while the S06 models do not take this into account.
The SED's for these objects, shown in Fig \ref{models}, indicate that the non-equilibrium are consistent with the
\lprime-M colors, while being discrepant with the [3.3]-\lprime\, colors.  Although the internal inconsistencies point toward a need for
better models and further observations, the current results all indicate the objects are
massive ($>$ 10 \mjup),  providing a challenge to models for explaining the stability of the system  \citep{Fabrycky09, Reidemeister09}, 
as well as its formation.  
 
\section{Summary}

Measurements of HR 8799 in thermal IR have constrained the SED of
these planets, helping to better understand their physical parameters.
The \lprime-M and [3.3]-\lprime colors of the objects appears to be blue, relative
to equilibrium models.  Models which account for non-equilibrium chemistry of CO
and CH$_4$ can account for this effect, but the details of the modeling leads to
differing results, depending on the input assumptions.  Improved modeling of these objects will help
understand the source of this.  Deeper 3.3\,\micron\, and M band photometry will
be helpful in constraining the SEDs of these interesting objects.

The effective temperatures of $b$, $c$, and $d$, derived from the colors as
predicted by the S06 models are $<$940, 1300, and 1170 K, indicating the
objects are $<$ 9 \mjup\, for $b$ and 12, and 11\mjup\, for $c$ and $d$ for an assumed age of 60 Myr. These estimates provide an
independent constraint  from the \citet{Marois08} results on the effective temperatures and masses of the companions to HR 8799. The 
results confirm the massive nature of this planetary system, and highlight the need for more detailed models to fit the colors of these
objects. 

\acknowledgements

 We are grateful to the staff of the MMT, especially Ricardo Ortiz and Tim Pickering, 
who were instrumental in obtaining the data.  We thank the Hectospec queue observers 
for graciously allowing us to exchange time. We thank Faith Vilas and Peter Strittmatter, 
for their assistance in enabling these observations. Thank you to Didier Saumon, Ivan Hubeny, Adam Burrows, and 
Jonathan Fortney for providing model spectra for comparison with the observations.

\bibliography{hinz}{}
\bibliographystyle{apj}

\clearpage

\clearpage

\clearpage

\clearpage

\clearpage


\begin{deluxetable}{rrrr}
\tabletypesize{\scriptsize}
\tablecaption{Log of data acquisition for Clio observations of HR 8799
\label{table1}}
\tablewidth{0pt}
\tablehead{
\colhead{Date} 		& \colhead{Total Integration (s)}  	& \colhead{Filter} 	& \colhead {Total Rotation (degrees)}  
}
\startdata 
21 November  2008 	&	5694						&	\lprime\,		&			72.0 \\
21 November  2008 &	9600						&	Barr M		&			31.8 \\
9 January 2009	&	2907						& 	3.3 \micron	&			5.3	\\
12 September  2009 &	6780						&	3.3 \micron	&			30.2 \\

\enddata

\end{deluxetable}

\begin{deluxetable}{ccr}
\tabletypesize{\scriptsize}
\tablecaption{Adopted Photometric Values for HR8799
\label{par}
}
\tablewidth{0pt}
\tablehead{
\colhead{Band} & \colhead{Stellar Magnitude}  & \colhead{zeropoint (Jy)} 
}
\startdata 
3.3\,\micron	&	5.23	&	330		\\
\lprime\,		&	5.22	&	235		 \\
M			&	5.21	&	154		\\

\enddata

\end{deluxetable}

\begin{deluxetable}{llrrr}
\tabletypesize{\scriptsize}
\tablecaption{Photometric Results
\label{phot}
}
\tablewidth{0pt}
\tablehead{
\colhead{System} &\colhead{Band} & \colhead{b}  & \colhead{c} 	& \colhead {d} 
}
\startdata 
Apparent Vega Magnitude	&3.3\,\micron\tablenotemark{a}	&	$>$1	7.8			&	15.63$\pm$0.3		&	15.53$\pm$0.3		\\
						&\lprime\tablenotemark{b}	&	15.81$\pm$0.23	&	14.99$\pm$0.12	&	14.39$\pm$0.19	 		 \\
						&M\tablenotemark{c}		&	$>$14.7			&	$>$14.7			&	$>$14.7				\\

\hline 		\\

Absolute Vega Magnitude &M$_{3.3}$		&	$>$14.82			&	12.65$\pm$0.3			&	12.55$\pm$0.3				\\
					&M$_{L'}$			&	12.83$\pm$0.23	&	12.01$\pm$0.12		&	11.41$\pm$0.3		 		 \\
					&M$_M$			&	$>$11.72			&	$>$11.72				&	$>$11.72				\\

\hline	\\	

Flux at 10 pc	 &F$_{3.3}$ (mJy)			&	$<$0.39			&	2.87$\pm$0.86			&	3.15$\pm$.95				\\
			&F$_{L'}$  (mJy)			&	1.73$\pm$0.41		&	3.69$\pm$0.44			&	6.41$\pm$1.9	 		 \\
			&F$_M$     (mJy)			&	$<$3.3			&	$<$3.3				&	$<$3.3				\\

\enddata
\tablenotetext{a}{Custom filter with zeropoint listed in Table \ref{par}.}
\tablenotetext{b}{MKO photometric system \citep{Tokunaga05}.}
\tablenotetext{c}{Bessell and Brett photometric system \citep{Bessell88}.}
\end{deluxetable}

\begin{deluxetable}{ccrrr}
\tabletypesize{\scriptsize}
\tablecaption{Astrometric Results for companions to HR 8799
\label{ast}
}
\tablewidth{0pt}
\tablehead{
\colhead{Date} 		& \colhead{Band} 	& \colhead{HR 8799 b}  			& \colhead{HR 8799 c} 			& \colhead {HR 8799 d } \\
\colhead{}			&\colhead{}		&\colhead{E [$\arcsec$], N [$\arcsec$]\tablenotemark{a}}  	&\colhead{E[$\arcsec$],N[$\arcsec$]\tablenotemark{a}} 	&\colhead{E[$\arcsec$],N[$\arcsec$]\tablenotemark{a}}
}
\startdata 
21 November 2008	&	\lprime		&	1.542$\pm$0.01, 0.780$\pm$0.014	&	-0.631$\pm$0.015,  0.671"$\pm$0.02	&	-0.215$\pm$0.021, -0.644$\pm$0.013 \\
9 January 2009	&	3.3\,\micron	&	--							&	-0.612$\pm$0.03, 0.665"$\pm$0.03		&	--	\\
12 September 2009	&	3.3\,\micron	&	--							&	-0.625$\pm$0.02, 0.725"$\pm$0.02		&	-0.282$\pm$0.03, -0.590$\pm$0.03 \\ 

\enddata
\tablenotetext{a}{Offset from HR8799.}
\end{deluxetable}
\clearpage





\end{document}